\documentclass[twocolumn]{aastex701}

\usepackage{amsmath} 
\usepackage{CJK}
\usepackage{hyperref}
\usepackage{booktabs}

\accepted{for publication in PNAS}

\begin{document}
\begin{CJK*}{UTF8}{gkai}

\title{Nonuniform Water Distribution in Jupiter's Mid-Latitudes: Influence of Precipitation and Planetary Rotation}

\author[orcid=0000-0001-6719-0759]{Huazhi Ge (葛华志)}
\affiliation{Division of Geological and Planetary Sciences, California Institute of Technology, 1200 E California Blvd, Pasadena, CA, 91125, USA}
\email[show]{huazhige@caltech.edu}  

\author[orcid=0000-0000-0000-0002]{Cheng Li (李成)} 
\affiliation{Department of Climate and Space Sciences and Engineering, University of Michigan, Ann Arbor, MI, 48109, USA}
\email{chengcli@umich.edu}

\author[orcid=0000-0002-8706-6963]{Xi Zhang (张曦)}
\affiliation{Department of Earth and Planetary Science, University of California Santa Cruz, Santa Cruz, CA, 95064, USA}
\email{xiz@ucsc.edu}

\author[orcid=0000-0002-2035-9198]{Andrew P. Ingersoll}
\affiliation{Division of Geological and Planetary Sciences, California Institute of Technology, 1200 E California Blvd, Pasadena, CA, 91125, USA}
\email[]{}

\author[orcid=0000-0002-0901-3428]{Sihe Chen (陈斯赫)}
\affiliation{Division of Geological and Planetary Sciences, California Institute of Technology, 1200 E California Blvd, Pasadena, CA, 91125, USA}
\email[]{}

\begin{abstract}

Knowing the composition of Jupiter's atmosphere is crucial for constraining Jupiter's bulk metallicity and formation history. Yet, constraining Jupiter's atmospheric water abundance is challenging due to its potential non-uniform distribution. Here, we explicitly resolve the water hydrological cycle in Jupiter's mid-latitudes using high-resolution simulations. Falling precipitation leads to a significant large-scale depletion of water vapor beneath the lifting condensation level. A non-uniform water vapor distribution emerges in the mid-latitude simulation with a changing Coriolis parameter across latitudes and spatially uniform cooling and heating. Water abundance at the 7-bar level varies by up to a factor of ten across latitudes, from sub-solar to super-solar values. We propose that nonlinear large-scale eddies and waves tend to drift air parcels across latitudes along constant potential vorticity (PV) surfaces, thereby sustaining latitudinal dependencies in water vapor and the interplay between water distribution and large-scale dynamics. Therefore, water distribution is influenced by the vertical structure of density stratification and changing Coriolis parameter across Jupiter's mid-latitudes, as quantified by PV. Additionally, the water hydrological cycle amplifies the specific energy of air parcels through the latent heat effect, thereby slowing down vertical mixing with a latent heat flux. The horizontal gradient of water is expected to be more pronounced with a super-solar water abundance. We suggest that similar interplays between precipitating condensates, planetary rotation, and distribution of condensable species generally exist in the weather layer of fast-rotating giant planets. The ongoing Juno mission and future Uranus mission may further reveal the non-uniform distribution of condensed species and their interplay with large-scale dynamics.
\end{abstract}



\section{significance statement} 
Jupiter is known for its colorful and dynamic appearance. However, its beauty poses a challenge for measuring its composition, as the optimal location for determining metallicity in Jupiter is largely unknown. Recently, Juno found that non-uniform features may extend to the layers well beneath clouds. Here, we examine water distribution in Jupiter's mid-latitudes. High-resolution simulations reveal a non-uniform distribution of water in Jupiter's weather layer. Precipitation establishes a large-scale depletion of water vapor tens of kilometers beneath water clouds. Turbulent large-scale eddies and waves lead to a latitudinal dependency of water vapor within the depleted levels. Our study highlights the significance of precipitation and large-scale dynamics in accurately measuring the bulk composition of condensable volatiles on giant planets.

\section{Introduction}

Jupiter accreted its materials from the protoplanetary disk, and its bulk metallicity provides crucial insights into the formation history of the solar system \citep{stevenson1988rapid,oberg2011effects}. However, the abundance of water vapor beneath its condensation level, a key parameter for constraining Jupiter's oxygen abundance and formation mechanism, remains highly uncertain due to the condensation of water vapor and potential spatial inhomogeneity driven by local weather or large-scale atmospheric dynamics \citep{showman1998interpretation}. The Galileo and Juno missions, designed to measure Jupiter's bulk water abundance, have conducted water vapor measurements at specific locations or latitudes near the equator \citep{niemann1996galileo,atreya1999comparison,wong2004updated,li2020water,li2024super}. Whether these measurements can accurately represent the bulk oxygen abundance of Jupiter's atmosphere remains unclear. Future measurements of heavy element abundance in ice giant atmospheres may encounter similar challenges as methane (the carrier of carbon) and hydrogen sulfide (the carrier of sulfur) condense in the cold atmospheres of these planets. Motivated by the importance of understanding the distribution of condensable species in giant planet atmospheres, we investigate the water distribution in Jupiter's atmosphere and identify the dominant physical processes influencing it.

The general circulation of the atmosphere modulates the distribution of chemical tracers in Jupiter's atmosphere \citep{conrath1984global,showman1998interpretation,ingersoll2017implications,zhang2018global,fletcher2020well,duer2021evidence,ingersoll2021jupiter}. Water, however, may act uniquely as an active chemical tracer by providing dynamical feedback on atmospheric circulations. Recent Juno observations found a reversed meridional brightness temperature gradient near the water condensation level, called the `jovicline' \citep{fletcher2021jupiter}. Hence, this finding adds evidence on the speculation that water vapor plays a unique role in Jupiter's atmospheric dynamics perhaps because water vapor carries significant latent heat and water vapor is substantially heavier than the background hydrogen-helium mixture \citep{gierasch1976jovian,guillot1995condensation,ingersoll2000moist,lian2010generation,li2015moist,leconte2017condensation,friedson2017inhibition,young2019simulatingII}. These effects combined may lead to a stratified layer with weak vertical motions near the condensation level of water \citep{sugiyama2011intermittent,sugiyama2014numerical,li2019simulating}. Studies on the Earth atmosphere have shown that the non-local condensation and settling of precipitation could remove water vapor beneath the LCL \citep{o2006stochastic,pierrehumbert2007relative,ding2016convection,pierrehumbert2016dynamics}. Furthermore, the large-scale distributions of ozone and radioactive chemicals in Earth's stratosphere are inhomogeneous and strongly influenced by the planetary rotation \citep{danielsen1968stratospheric,haynes1990conservation,holton2004introduction}. The potential vorticity (PV), being a nearly conserved quantity in isentropic motions, tracks the changes in the swirl of the fluid in
response to planetary rotation and reveals information about the flow and dynamics \citep{hoskins1985use,haynes1990conservation,mcintyre2003potential}. Nevertheless, the significance of PV in diagnosing chemical tracer distribution was rarely applied for composition measurements in planetary atmospheres \citep{showman1998interpretation,zuchowski2009modeling,o2016slantwise}. In this work, we demonstrate that analyzing precipitation and quasi-conserved tracers, including PV, is essential for understanding the distribution of water vapor in Jupiter's mid-latitude atmosphere, which has remained largely unexplored.

Our study is motivated by three questions: (I) How does the water vapor distribute beneath its lifting condensation level? (II) What influences the water vapor distribution if it is not homogeneous? (III) How can we improve our understanding of chemical distributions in planetary atmospheres, especially those that matter to measuring metallicity? We will first present our findings from the numerical simulation and then provide explanations to address the listed questions.

\section{The Simulated Water Distribution}
\label{sec:water-distribution}

\begin{figure*}[t]
\centering
\includegraphics[width=17.8cm]{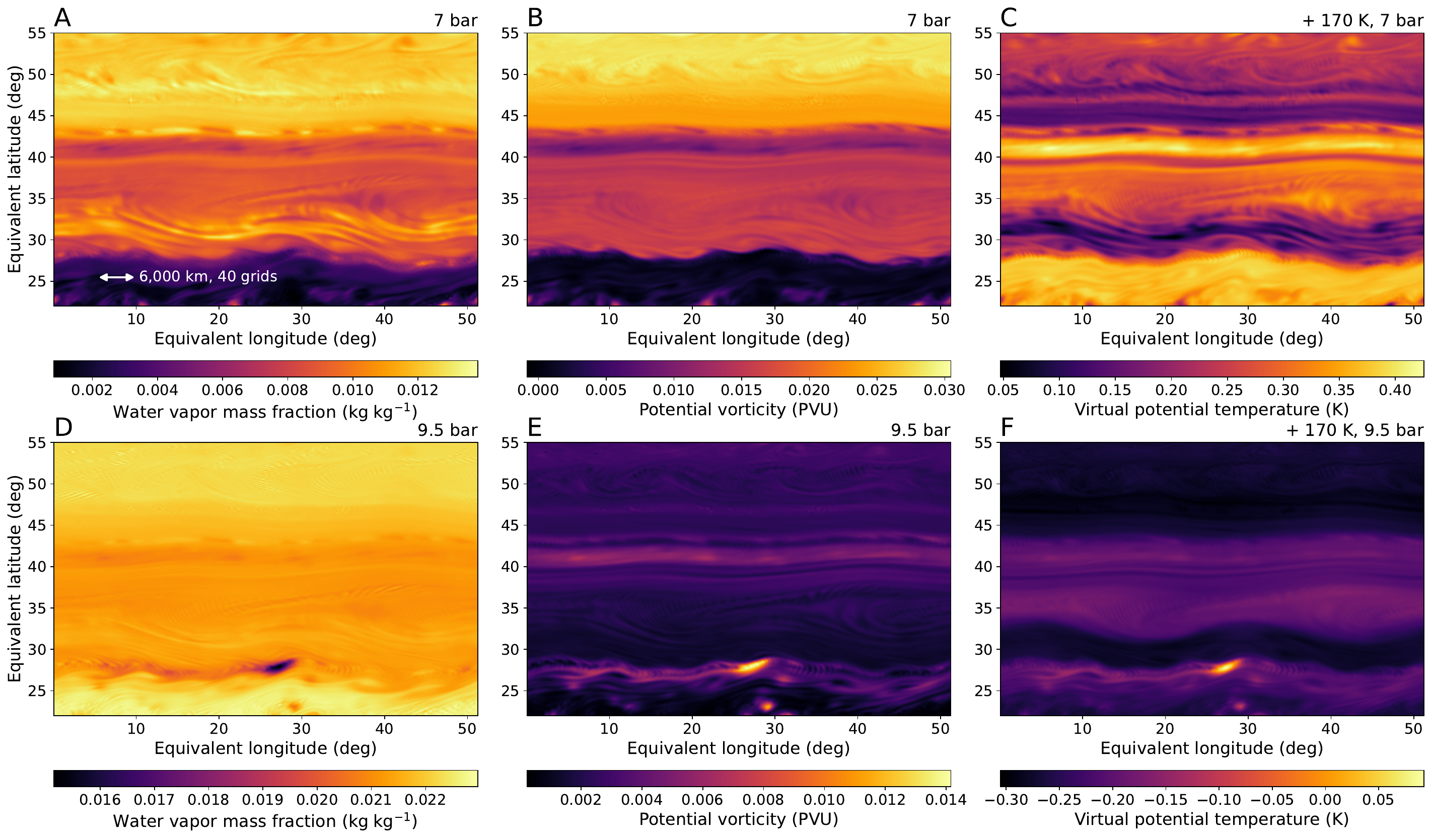}
\caption{A snapshot of water vapor mass fraction (A and D), potential vorticity (B and E), and virtual potential temperature (C and F) at 7-bar level (A-C) and 9.5-bar level (D-F) on simulation Day 4,180. As a reference, the LCL of water vapor is at 5.5 bar, and the average 1-bar temperature is about 172 K (\textit{SI Appendix} Fig.~1). Notably, the color bars of the same quantity have different color scales at different pressure levels. The white arrow in panel A represents a horizontal distance of 6,000 km, resolved by 40 grid points in our model. At the 9.5-bar level, the water vapor map is anti-correlated with the PV map but correlated with the virtual potential temperature map. The equivalent latitude $\phi$ is computed by the gradient of the planetary vorticity $\beta_{0}$ and the local Coriolis parameter $f = f_{0} + \beta_{0}y = 2\Omega\sin{\phi}$, where $f_{0}$ is the Coriolis parameter at $20^\circ$ N and $y$ is the meridional distance to the southern boundary. The equivalent longitude $\lambda$ is calculated by $\lambda = 180^\circ\arcsin{(x/\pi R_{p})}$ where $x$ is the horizontal distance and $R_{p}$ is Jupiter's radius.}
\label{fig:fig1-water}
\end{figure*}

We simulated the water vapor distribution in Jupiter's mid-latitudes using the nonhydrostatic model SNAP \citep{stone2020athena++,li2019simulating,ge2020global} in a Cartesian box with linearly varying Coriolis parameter across the latitude (called beta-plane approximation). The simulation domain spans 45,000 km in latitude (y-direction) and 60,000 km in longitude (x-direction). We convert the distances in $x-$ and $y-$directions to equivalent longitude and latitude with the radius of Jupiter and local Coriolis parameter on the $\beta$ plane (see Fig.~\ref{fig:fig1-water} caption). Vertically, the domain extends from $\sim$87 bar (-240 km) at the lower boundary to $\sim$0.002 bar (100 km) at the top, with the reference height ($z_{0} = 0$ km) set at 1 bar. The spatial resolution is 150 km in both longitude and latitude (about $0.1^{\circ}$) and 2 km in the vertical direction. The spatial resolution allows us to resolve the Rossby deformation radius and the scale height of water vapor. The initial water vapor abundance in the deep reservoir is assumed to be three times solar and horizontally uniform. We applied spatially uniform cooling at the top and heating at the bottom without imposing differential solar heating from above. The simulation starts without prescribed background flows, and reaches a statistically steady state after approximately eight Earth years (\textit{SI Appendix} Fig.~1 and \textit{SI Appendix} Movie~2). Additional details about the numerical simulation can be found in \textit{Materials and Methods}.

\begin{figure}[tb]
\centering
\includegraphics[width=\linewidth]{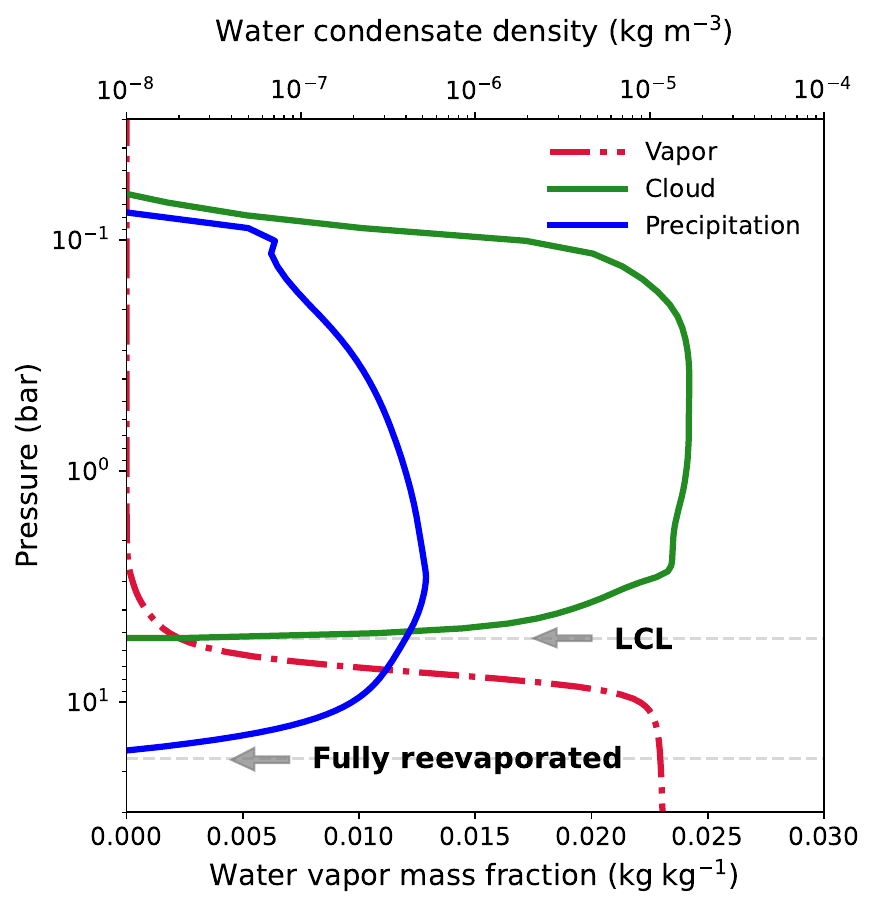}
\caption{Horizontally and temporally averaged vertical profiles of the simulated water vapor mass fraction (red dashed line), water cloud density (green solid line), and water precipitation density (blue solid line) averaged from the $7_{\rm th}$ to $15_{\rm th}$ simulation year. The bottom abscissa measures the vapor mass fraction, while the top abscissa measures the condensate density. Simulation indicates that water vapor is depleted from the LCL at about 5.5 bar to the level where the water precipitation fully reevaporates. The mean precipitation flux is about $\sim\mathcal{O}(100\;\rm mm\;yr^{-1})$.}
\label{fig:fig2-hydrological-cycle}
\end{figure}

\begin{figure*}[tb]
\centering
\includegraphics[width=17.8cm]{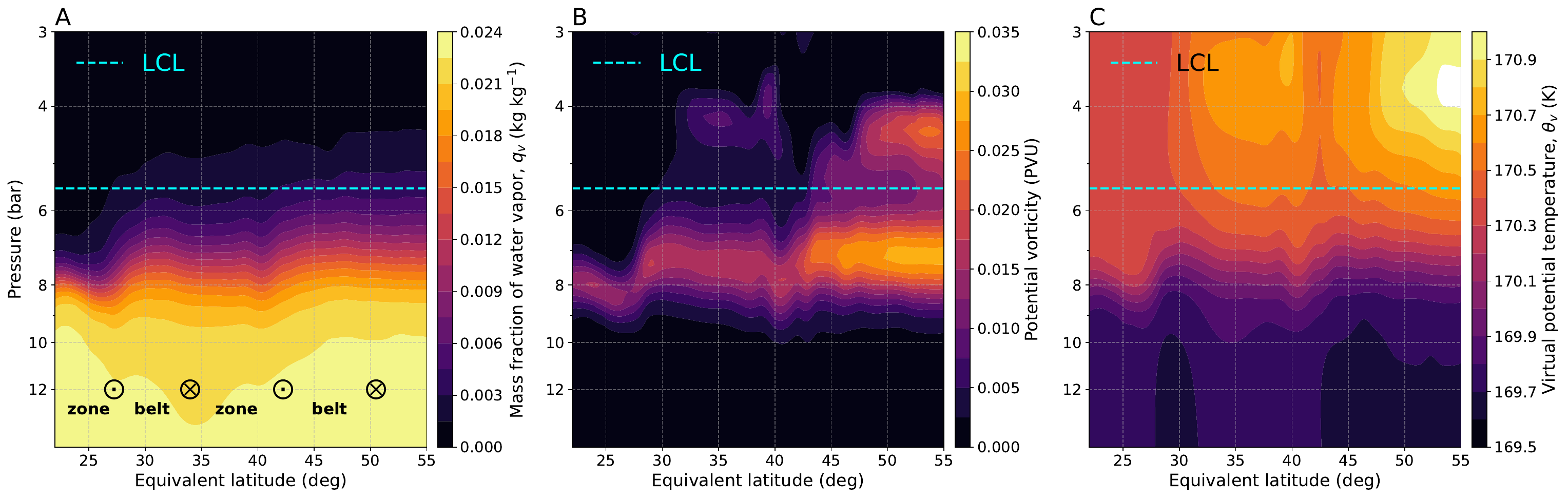}
\caption{Panels A-C respectively show the 2D distribution of temporally and zonally averaged water vapor mass fraction $q_{v}$, PV, and virtual potential temperature $\theta_{v}$ in the latitude-pressure cross-section from Day 4,170 to Day 4,300. We first calculate PV using the local density, buoyancy frequency, and the Coriolis parameter, and then take the average. In panel A, circled dots denote eastward prograde jets, while the circled crosses denote westward retrograde jets.}
\label{fig:fig2-PV-water-thetav}
\end{figure*}

Fig.~\ref{fig:fig1-water}A and D show a snapshot of the water vapor mass fraction (i.e., specific humidity) at two constant pressure surfaces in longitude and latitude. Fig.~\ref{fig:fig2-hydrological-cycle} shows the horizontally averaged profiles of water vapor and condensates and Fig.~\ref{fig:fig2-PV-water-thetav}A shows the cross-section of zonally and temporally averaged water vapor distribution in latitudes and pressure from 15 bars to 3 bars. Fig.~\ref{fig:fig1-water}A and D and Fig.~\ref{fig:fig2-hydrological-cycle} indicate that the simulated distribution of water vapor has a significant vertical depletion beneath the lifting condensation level (LCL) down to approximately 15 bars across the entire mid-latitude region (Fig.~\ref{fig:fig1-water}A). While the deep water abundance is 0.023 $\rm kg\;kg^{-1}$ ($\sim$ 3 $\times$ Solar), water vapor abundance at the 7-bar level varies from about 0.001 $\rm kg\;kg^{-1}$ ($\sim$ 0.15 $\times$ Solar) to 0.014 $\rm kg\;kg^{-1}$ ($\sim$ 2 $\times$ Solar). At the 9.5-bar level, the abundance of water vapor remains depleted relative to the deep reservoir. Water vapor becomes nearly well-mixed at about 15 bars, where most precipitation reevaporates (Fig.~\ref{fig:fig2-hydrological-cycle}). Unlike the equilibrium condensation cloud models showing well-mixed vapor right beneath the LCL \citep{weidenschilling1973atmospheric}, the depletion of water vapor in our simulation is more consistent with the advection-condensation model from the Earth studies, which suggests that non-local condensation can remove water from air parcels \citep{o2006stochastic,pierrehumbert2007relative,sukhatme2011advection}. Non-local condensation could manufacture dry air beneath the LCL by advecting a wet air parcel from the deep atmosphere to a cold-enough region, where water may condense and precipitate out, and then bringing the dehydrated air parcel back to the deep atmosphere, for example, by large-scale eddies \citep{pierrehumbert1993global,yang1994production} or waves \citep{showman1998interpretation,showman2000nonlinear}, which is generally more consistent with our simulation result and in-situ measurements from the Galileo probe showing that the 5 $\rm \mu$m hot spot near $\rm 7^\circ$ N has a local depletion of water vapor beneath LCL \citep{niemann1996galileo,wong2004updated}. Our simulation suggests that water vapor depletion generally occurs across all latitudes and longitudes and is related to precipitation (Fig.~\ref{fig:fig2-hydrological-cycle}). In the later section, we will further explain how falling precipitation establishes large-scale water vapor depletion.

In addition to the vertical depletion of water vapor across the domain, one prominent finding in our simulation is that the water vapor varies significantly with latitude on constant pressure surfaces (Fig.~\ref{fig:fig2-PV-water-thetav}A), without external radiative forcing. At the 7-bar level, the difference in the water vapor abundance could be a factor of more than ten (Fig.~\ref{fig:fig1-water}A). Water vapor maps in Fig.~\ref{fig:fig1-water}A and D also reflect the dynamical footprints of various types of longitudinally circulating bands and waves. Belts, characterized by cyclonic flows, contain slightly more water than anticyclonic zones at the 7-bar level, but the belt/zone difference is less significant than the general latitudinal difference across the mid-latitude. Besides, water is slightly depleted than neighboring latitudes within eastward prograde jets at about $\sim 27^{\circ}$ N and $\sim 42^{\circ}$ N (Fig.~\ref{fig:fig1-water}A). Turbulent wave features on the water vapor map indicate that large-scale dynamics influence tracer distribution a few scale heights below the superficial clouds. Although the simulation domain does not cover 5 $\rm \mu m$ hot spots at $\rm \sim 9^{\circ}$ N, the general trend set by the meridional gradient is consistent with the severe depletion of water vapor in these meteorologically anomalous regions.

\section{Vertical Depletion of Water Vapor and Stratification Maxima Formed by Falling Precipitation}
\label{sec:water-depletion-stratification}

We find that falling precipitation sustains the large-scale depletion of water vapor from the LCL to the level where precipitation fully evaporates. At the statistically steady state, the large-scale vertical depletion of water vapor can be viewed and simplified as a 1D problem that only concerns altitude $z$ by taking the horizontal average to remove latitudinally or longitudinally dependent features (Fig.~\ref{fig:fig2-hydrological-cycle}). The bottom line of the water hydrological cycle is that the net vertical water vapor flux must be balanced by a recycling flux that can bring water downward. The downward flux can occur in a condensed phase, such as liquid or ice particles. Condensed particles may decouple from the background atmospheric motions if they grow large enough. Hence, falling precipitation is suitable for closing the hydrological cycle by recycling water back into the deep reservoir.

We could quantify this process by analyzing the mass conservation of total water and total air. We here employ Reynolds' decomposition to analyze the hydrological cycle of water at the steady state, for an arbitrary quantity $A$, $A = \overline{A}(z) + A'(x,y,z,t)$ (see derivations in Eq.~\ref{eq:total-water-continuity}-\ref{app-eq:water-eddy-transport}). The balance of water fluxes in vapor and condensate reads as:
\begin{equation}
\label{eq:eddy-water-flux}
    \underbrace{\overline{q_{v}'w'}}_{ \substack{\text{vapor} \\ \text{flux}} } = -\dfrac{(1 - \overline{q_{v}})}{\overline{\rho}}\underbrace{\overline{\rho_{p}w_{T}}}_{ \substack{\text{precipitation} \\ \text{flux}} } > 0,
\end{equation}
where $q_{v} \sim \mathcal{O}(0.01\;{\rm kg/kg}) \ll 1$ on Jupiter, $w$ is the vertical velocity, $\rho$ is the total density of the air parcel including materials in all phases, $\rho_{p}$ is the density of precipitation, $w_{T}$ is the vertical terminal velocity of precipitation that is always negative and point to the deep atmosphere since gravity pull condensates downward.

Eq.~\ref{eq:eddy-water-flux} informs that the net eddy water vapor flux essentially results from the upward advection of water-enriched and the downward advection of water-depleted air parcels at small scales. Note that the water-enriched air parcels with positive water vapor anomaly $q_{v}' > 0$ are created by reevaporation of precipitation. They are naturally resistant to moving upward since they are usually thermally colder than the background with a larger molecular weight due to the reevaporation. 

If we approximate the net eddy water vapor flux as a down-gradient turbulent diffusive process with a positive eddy diffusivity $\mathcal{K}_{q_v}$ \citep{vallis2017atmospheric}. Then, the averaged vertical gradient of water vapor shall be negative unless precipitating condensates are fully reevaporated or, equivalently, the precipitation flux attenuates to zero
\begin{equation}
\label{eq:water-kzz}
\begin{split}
    \overline{q_{v}'w'} & = -\mathcal{K}_{q_v}\overline{\frac{\partial q_{v}}{\partial z}}\Big\vert_{p > {\rm LCL}} = -\dfrac{(1 - \overline{q_{v}})}{\overline{\rho}}\overline{\rho_{p}w_{T}} > 0, \\
    & \Rightarrow \; \overline{\frac{\partial q_{v}}{\partial z}}\Big\vert_{p > {\rm LCL}} < 0.
\end{split}
\end{equation}
This approach assumes a flux-gradient relation from the conservation of total water to simplify the 3D hydrodynamic simulation. We should note that Eq.~\ref{eq:water-kzz} and the use of eddy diffusivity may not capture the non-local removal of water vapor that account for the generation of the precipitation, which is a consequence of the non-linear nature of the phase transition when condensation occurs, especially if the lifting condensation level has a significant meridional change due to the meridional temperature contrast \citep{o2006stochastic,pierrehumbert2007relative,sukhatme2011advection}. Nevertheless, our simulation directly resolves these non-local processes and adds the evidence that precipitation can remove water vapor beneath the LCL (Fig.~\ref{fig:fig2-hydrological-cycle}). 

The simulated water depletion extends to about 15 bars with a fixed e-folding reevaporation timescale, $t_{\rm eva} = 3.3\times 10^{2}$ s, and terminal velocity, $w_{T} = -10\;{\rm m\;s^{-1}}$. These microphysics parameters, which relate to the depth of water depletion, are chosen according to studies that suggest raindrops can fall approximately one scale height after being detached from the cloud level \citep{ohno2017condensation,loftus2021physics}. Larger ice condensates like hailstone or mushball may fall and evaporate deeper to tens of bars on Jupiter \citep{guillot2020stormsI,markham2023rainy}, which may lead to a deeper depletion of water vapor on Jupiter. If we utilize a probe to measure Jupiter's deep water vapor abundance or any other condensable species on giant planets, we may need to know the depth to which precipitation can survive and reach, so the probe can reach that level.

Despite the vertical depletion of water vapor, falling precipitation causes a side effect in the atmosphere by forming a stratification maxima at the $\sim$7.5-bar level in Fig.~\ref{fig:Fig3-sketch}B. Previous studies have reported similar findings \citep{sugiyama2011intermittent,sugiyama2014numerical,li2019simulating,ge2024heat}, and we provide a quantitative estimation here. The falling precipitation experiences an upward drag force balanced by gravity. Conversely, due to Newton's third law, the background atmosphere experiences a downward drag force that equivalently enhances gravity. The oscillation of the drag force between the rainfall and no-rainfall states eventually leads to a synoptic-scale, long-lived stratified layer beneath the condensation level. We can estimate the horizontally averaged buoyancy frequency at the stratification maxima from the budget of (I) the net buoyancy flux and (II) the net mass flux of total water (see detailed derivation in Eq.~\ref{app-eq:buoyancy-flx}-\ref{app-eq:stratification}),
\begin{equation}
\label{eq:stratification}
\begin{split}
    \overline{N^{2}}  & = \overline{\frac{g}{\theta_{v}}\frac{\partial \theta_{v}}{\partial z}} \approx -g\overline{\frac{\partial q_{v}}{\partial z}} \approx g\frac{q_{v,d}}{H_\text{eva}} \sim \mathcal{O}(10^{-5} \; {\rm s^{-2}}),
\end{split}
\end{equation}
where $H_{\rm eva}$ is a characteristic distance from the level that precipitation can fall from the cloud base (the simulated $H_\text{eva}$ is about one scale height), and $q_{v,d}$ is the water vapor mass fraction in the deep reservoir which is about 0.023 $\rm kg\;kg^{-1}$ for 3 times solar of oxygen. The estimated buoyancy frequency is qualitatively consistent with the simulated number in Fig.~\ref{fig:Fig3-sketch}B. The buoyancy frequency induced by precipitation would attenuate to zero in both deep reservoir and upper troposphere where both the precipitation flux and the vertical gradient of water vapor are nearly zero, $\partial_{z}q_{v} \rightarrow 0$ (Fig.~\ref{fig:Fig3-sketch}B).

\section{Influence of Jupiter's Rotation: Water Vapor Distribution and Potential Vorticity}
\label{sec:PV-water}

Precipitation sets the vertical depletion of water vapor in the averaged 1D scope. However, if we wish to measure the metallicity of the entire planet, we shall also consider the longitudinal and latitudinal distributions of water vapor beyond the depth of vertical depletion, as these measurements may be biased in specific locations at certain levels. Jupiter's mid-latitudes are expected to be more homogeneous in longitude than in latitude due to zonal jets. How do 1D columns in Fig.~\ref{fig:fig2-hydrological-cycle} interact and sustain the inhomogeneity of water vapor in Fig.~\ref{fig:fig1-water} and~\ref{fig:fig2-PV-water-thetav}A?

\begin{figure*}[t]
\centering
\includegraphics[width=17.8cm]{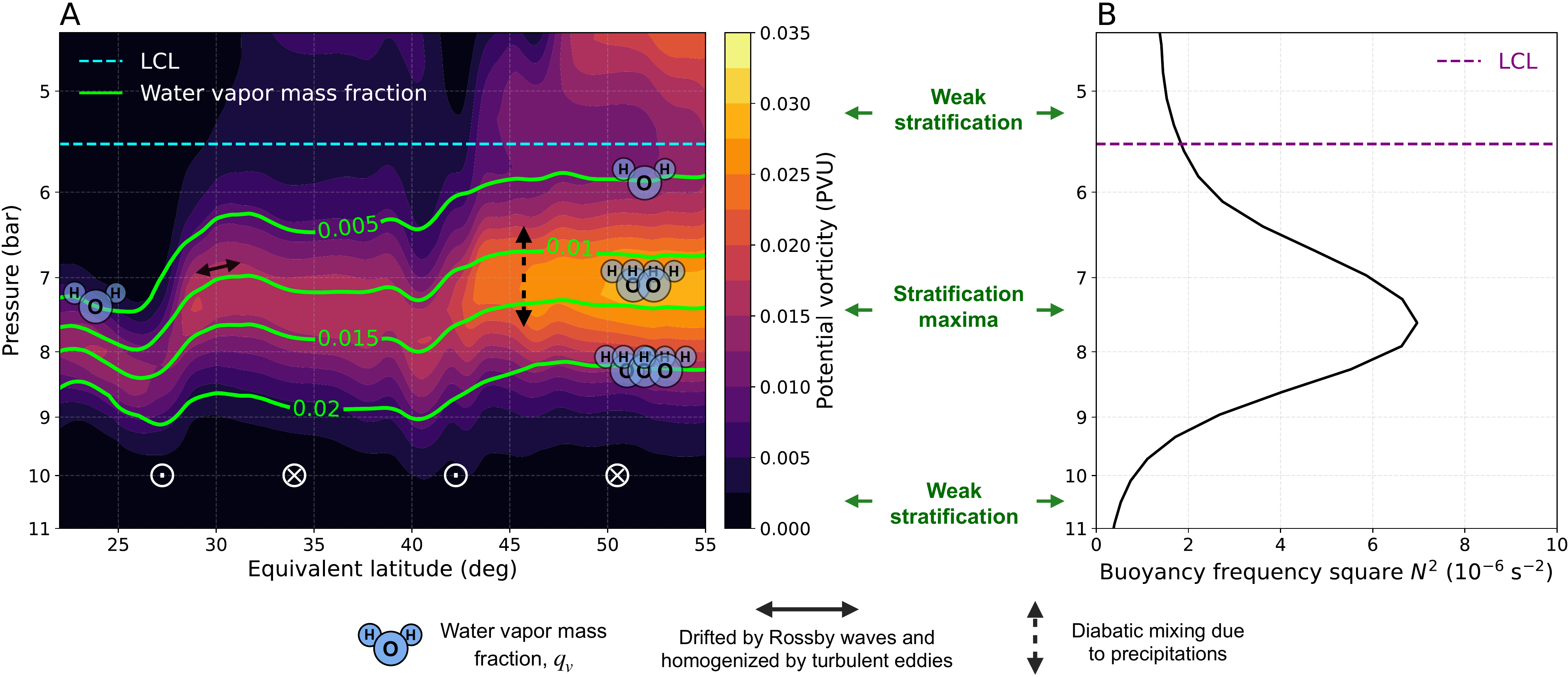}
\caption{Panel A shows the zonally and temporally averaged water vapor mass fraction (green solid isolines) and PV (the color-filled contour map) from Day 4,170 to Day 4,300. Panel B shows the horizontally and temporally averaged buoyancy frequency calculated from the vertical gradient of $\theta_{v}$. Panel B shows the stratification peaks at about 7.5 bar. We propose that nonlinear Rossby waves tend to drift air parcels along PV isolines, and turbulent mixing may occur to homogenize water vapor along the trajectory (black solid arrow), thereby sustaining a general correlation between PV and water vapor. On the other hand, diabatic mixing driven by precipitation mostly occurs in the vertical direction, as the falling precipitation responds to gravity (black dashed arrow). PV isolines in panel A are mostly determined by the absolute vorticity $f+\zeta$ and stratification $N^2$. $f$ changes by a factor of three from $\rm 20^\circ$ to $\rm 55^\circ$ in latitudes, and $N^2$ could change by more than a factor of ten from about $7\times 10^{-6}\;{\rm s^{-2}}$ to much less than $10^{-6}\;{\rm s^{-2}}$ in the vertical. In contrast, density $\rho$ changes from 0.52 $\rm kg\;m^{-3}$ to 0.75 $\rm kg\;m^{-3}$ (less than 50\%), and virtual potential temperature $\theta_{v}$ changes from $\sim 170.5$ K to $\sim 169.5$ K ($\sim$1\%).}
\label{fig:Fig3-sketch}
\end{figure*}

We propose that nonlinear large-scale eddies and waves which can propagate within the stratified atmosphere with a changing Coriolis parameter across latitudes (called the $\beta$ effect by meteorologists) sustain the non-uniform distribution of water vapor shown in Fig.~\ref{fig:fig1-water}A and D,~\ref{fig:fig2-PV-water-thetav}A, and~\ref{fig:Fig3-sketch}A. In our simulations, the latitudinal gradient of water vapor initially forms from a large-scale convective mixing during the spin-up phase (\textit{SI Appendix} Movie 2). Subsequently, large-scale eddies and waves, such as baroclinic eddies and edge waves propagating near the prograde jets (hereafter turbulent Rossby waves), sustain these large-scale structures by drifting air parcels in latitudes (\textit{SI Appendix} Fig.~4 and 5). 

We employ potential vorticity (PV) to analyze the system. PV describes a nearly conserved scalar field in isentropic motions that can roughly reveal the trajectory of air parcels drifted by turbulent Rossby waves. It mostly concerns the planet's rotation quantified by the Coriolis parameter $f$, fluid spinning quantified by the relative vorticity $\nabla\times\boldsymbol{u}$, and atmosphere stratification quantified by the square of Brunt-V\"ais\"al\"a buoyancy frequency $N^2$. The definition of PV in the context of Jupiter's atmosphere could be,
\begin{equation}
\label{eq:PV-def}
\begin{split}
    {\rm PV} & = \underbrace{(2 \boldsymbol{\Omega}+\nabla\times\boldsymbol{u})}_{\substack{\text{absolute } \text{vorticity}}} \cdot \underbrace{\frac{\nabla\theta_{v}}{\rho}}_{\substack{\text{weighted} \\ \text{stratification}}} \approx (f + \zeta)\frac{1}{\rho}\frac{\partial \theta_{v}}{\partial z} \\ 
    & = (f + \zeta)\frac{\theta_{v}}{\rho g}N^2,
\end{split}  
\end{equation}
where $\boldsymbol{\Omega}$ is the angular velocity of the planet, $f = 2\Omega\sin{\phi} = f_{0} + \beta_{0} y$ is the Coriolis parameter that increases in latitude $\phi$, $\theta_{v}$ is the virtual potential temperature that represents the potential density of the air parcel adjusted by isentropic expansion or compression (i.e., buoyancy), $\zeta = (\nabla\times\boldsymbol{u})\cdot\hat{z}$ is the vertical component of relative vorticity, $g$ is the gravity. The approximation in Eq.~\ref{eq:PV-def} could represent more than 99\% of total PV, while the rest only contributes to less than 1\% (see \textit{SI Appendix} Fig.~6). 

\begin{figure*}[t]
\centering
\includegraphics[width=\linewidth]{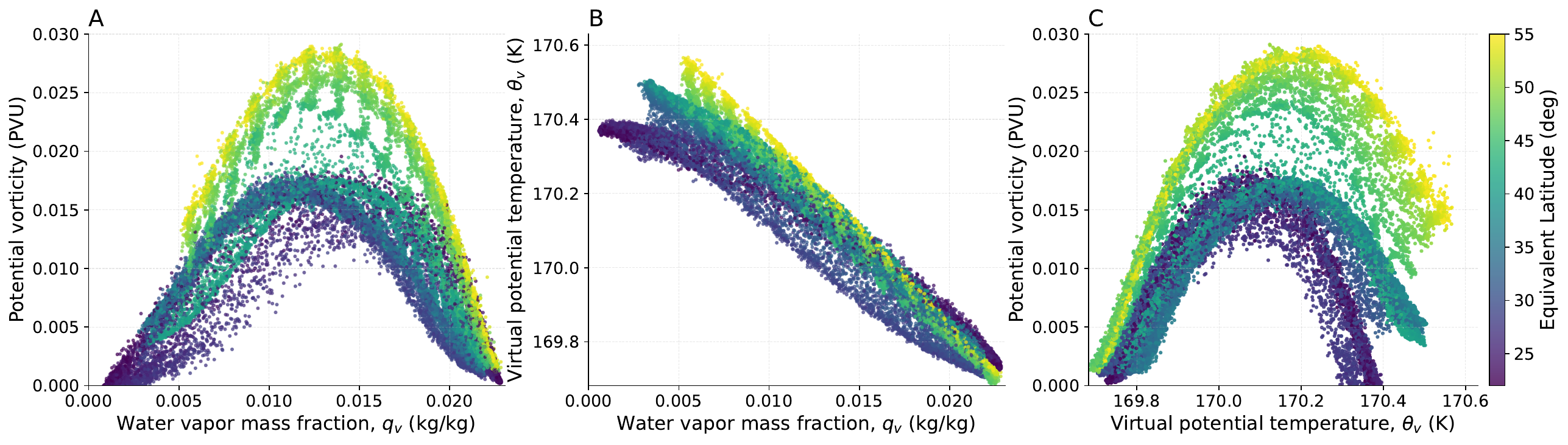}
\caption{Scattering plots of water vapor mass fraction $q_{v}$, PV, and virtual potential temperature $\theta_{v}$ colored by the equivalent latitude. These dots are collected from a snapshot in the simulation from 6 to 10 bars on Day 4,180 (same day as the data in Fig.~\ref{fig:fig1-water}). Panel A shows $q_{v}$ versus PV; panel B shows $q_{v}$ versus $\theta_{v}$; and panel C shows $\theta_{v}$ versus PV. Panels A and C show a general correlations between PV and other quasi-conserved tracers shift to general anti-correlations (at $q_{v} \sim 0.012 \;{\rm kg/kg}$ in panel A and $\theta_{v}\sim 170.1\;{\rm K}$ in panel C) because PV first increases in altitude and then decreases. Both the vertical variations of water vapor and virtual potential temperature are monotonic, while the vertical variation of PV is non-monotonic due to the stratification peak near the 7.5-bar level. All three panels show a separation of points characterized by three groups, mostly colored by cosmic cobalt, green, and yellow dots, which reflect the distribution of tracers within three staircases at 22$^\circ$-27$^\circ$ N, 27$^\circ$-42$^\circ$ N, and 42$^\circ$-55$^\circ$ N. The staircase separation and spreading of PV with a $q_{v}$ and $\theta_{v}$ indicates that staircase mixing is more efficient than global mixing. $q_{v}$ and PV distributions on constant $\theta_{v}$ maps are provided in \textit{SI Appendix} Fig.~3.}
\label{fig:Fig4-correlation}
\end{figure*}

We find the leading terms in Eq.~\ref{eq:PV-def} that govern the simulated PV structure in Fig.~\ref{fig:fig2-PV-water-thetav} and~\ref{fig:Fig3-sketch} are the Coriolis parameter $f$, which increases in latitude, and the stratification $N^2$, which is set by precipitation and peaks near 7.5 bars (Fig.~\ref{fig:Fig3-sketch}B). The variations of the other quantities are generally more minor than the change of $f$ and $N^2$ across the domain (see Fig.~\ref{fig:Fig3-sketch} caption for numbers). If PV is generally conserved along the drift trajectory, the path becomes a function of both altitude in $z$ and latitude in $y$, as the vertical change in stratification $N^2$ can offset the meridional change in $f$. Consequently, turbulent Rossby waves would drift air parcels along sloped trajectories that generate the non-uniform distribution of water vapor in the $y-z$ cross-section. We suggest that this sloped mixing sustains the latitudinal dependency of water vapor distribution in Fig.~\ref{fig:fig1-water}A and D. In turn, the redistribution of water vapor that deviate it from the 1D profile in Fig.~\ref{fig:fig2-hydrological-cycle} feeds back into the above picture by altering the buoyancy structure (i.e., $\theta_{v}$ distribution), as air parcels with more water vapor tend to possess larger potential density and a smaller virtual potential temperature (Fig.~\ref{fig:Fig4-correlation}B). Therefore, the redistributed water vapor would slightly alter the atmospheric stratification $N^2$ in the latitudinal direction. This feedback is resolved in our simulation, whereas the simulated $\theta_{v}$ structure is similar to the water vapor distribution (Fig.~\ref{fig:fig2-PV-water-thetav}A and C). 

The picture above qualitatively explains the enrichment of water vapor in higher latitudes at approximately the 7-bar level (Fig.~\ref{fig:fig1-water}A). A place with a small $f$ in low latitudes and a large buoyancy frequency, $N$, could have the same PV as another place in high latitudes with a large $f$ and a small $N$. Hence, at this level, turbulent Rossby waves would tend to drift and mix water-enriched air parcels poleward and upward, or, conversely, drift and mix water-depleted air parcels equatorward and downward. The general homogenization of water vapor along the PV isolines would minimize the angle between the PV and water vapor isolines, which sustains the non-uniform distribution of water vapor and a general enrichment of water vapor in higher latitudes at the 7-bar level (see \textit{SI Appendix} for further analysis on the turbulent mixing of water vapor). 

Although the simulated water vapor distribution is similar to the PV structure in Fig.~\ref{fig:fig1-water}, the isolines of water vapor and PV are not perfectly aligned. Simulation indicates that the homogenization of water vapor within each staircase is more efficient than the mixing at the global scale (Fig.~\ref{fig:Fig4-correlation} and \textit{SI Appendix} Fig.~3). PV is relatively flat on constant pressure surfaces within each staircase at 30$^\circ$ to 40$^\circ$ N and from 45$^\circ$ to 55$^\circ$ N as shown in Fig.~\ref{fig:fig1-water}B and~\ref{fig:Fig3-sketch}A. The jet formation slightly complicates the above picture, as the meridional change of $f$ is partially canceled by the relative vorticity $\zeta$ of jets within staircases. The sloped structures are more evident in regions near prograde jets, where the meridional change of absolute vorticity $f+\zeta$ in Eq.~\ref{eq:PV-def} is significant. However, the discussion and analysis of the formation of jets and possible baroclinic eddies are beyond the scope of this study.

Falling precipitation is a competitive process to nonlinear mixing related to eddies and nonlinear waves. Falling precipitation is necessary to sustain the vertical depletion of water vapor and maintain its inhomogeneity. However, diabatic processes associated with the reevaporation of settled precipitation can disrupt the large-scale patterns and force air parcels to move across the isolines of PV or $\theta_{v}$ before the waves drift them back if the cloud-level moist convection above the LCL is vigorous. Therefore, the diabatic mixing due to moist convection should be slow enough to maintain the formation of the horizontal structure that we see in Fig.~\ref{fig:fig1-water}. We will estimate and compare these timescales in the next section.

\section{On the Timescales of the Homogenization of Water Vapor and Diabatic Processes Related to Reevaporation of Precipitation}
\label{sec:conservation-dissipation}

\begin{table}
\centering
\caption{A table of typical timescales of different dynamical processes (from short to long) associated with the water system in Jupiter's mid-latitude if water abundance is three times solar.}
\begin{tabular}{lc}
\midrule
Physical processes & Timescales \\
\midrule
1. Vertical oscillations due to stratification$^{1}$ & $\mathcal{O}(10^3$ s) \\
2. Homogenization driven by Rossby wave $^{2}$ & $\mathcal{O}(10^6$ s) \\
3. Hydrological cycle (diabatic mixing)$^{3}$ & $\mathcal{O}(10^8$ s) \\
4. diabatic changing timescale of water vapor$^{4}$ & $\mathcal{O}(10^8$ s) \\
5. diabatic changing timescale of PV$^{5}$ & $\mathcal{O}(10^8$ s) \\
6. Radiative cooling$^{6}$ & $\mathcal{O}(10^8$ s) \\
7. diabatic changing timescale of buoyancy$^{7}$ & $\mathcal{O}(10^{10}$ s) \\

\bottomrule
\label{table:timescales}
\end{tabular}

\caption{References: 1. Fig.~\ref{fig:Fig3-sketch}B; 2. $\tau_{\rm R}$; 3. Table 1 in \citep{ge2024heat}; 4. Eq.~\ref{eq:tau-qv}; 5. Eq.~\ref{eq:tau-thetav}; 6. \citep{gierasch1976jovian,li2015moist}; 7. Eq.~\ref{eq:tau-PV}}.
\end{table}

Large-scale mixing and the reevaporation of precipitation are both diabatic. To distinguish between them, we refer to the former as the homogenization of water vapor along PV isolines and the latter as the diabatic change due to the reevaporation of precipitation.

It is challenging to estimate the timescale of large-scale mixing, as it strongly depends on nonlinear processes. Here, we assume that the limiting timescale for homogenization corresponds to the timescale of a typical Rossby wave cycle. The typical cycle of Rossby waves could be obtained from the Rhines scale $L_{\rm R} \sim ({v'/\beta_{0}})^{1/2}$ and a typical meridional velocity $v'$. The Rhines scale quantifies the distance at which air parcels can feel the changing Coriolis parameter when moving across latitudes, and it is generally a typical distance that the air parcel could be drifted meridionally. The typical Rhines scale is about $L_{\rm R} \sim ({v'/\beta_{0}})^{1/2} \sim \mathcal{O}(3,000$ km) with a typical eddy meridional velocity $v' \sim \mathcal{O}(3\;{\rm m\;s^{-1}})$ from the simulation. $v'$ in Jupiter's atmosphere could be further constrained by cloud-tracking observations \citep{tollefson2017changes}. The homogenization timescale is then $\tau_{H} \sim L_{\rm R}/v' \sim (\beta_{0}v')^{-1/2} \sim \mathcal{O}(10^6\;\text{s}) \sim$ ten days.

On the other hand, the diabatic changing timescale, $\tau_{A}$, of any scalar quantity $A$, could be inferred from its continuity equation if the equation could be written in the form of
\begin{equation}
    \frac{DA}{Dt} = \dot{A} = \frac{A}{\kappa} \; \Rightarrow \; \tau_{A} = \vert \kappa\vert = \Big\vert\frac{A}{\dot{A}}\Big\vert,
\end{equation}
We first estimate the timescale $\tau_{q_{v}}$ for water vapor, which relates to the hydrological timescale (see Eq.~\ref{eq:full-qv} and \ref{app-eq:tau-qv} for derivations). We anticipate that most of the diabatic processes in Jupiter's opaque troposphere will be determined by the latent heat and virtual effect resulting from the phase transition of water vapor. The mass flux of precipitation is constrained by the available heat flux for moist convection in the cloud layer above the LCL and specific latent heat \citep{guillot2020stormsII,ge2024heat},
\begin{equation}
\label{eq:tau-qv}
\begin{split}
    \tau_{q_{v}} & = \Big\vert -\frac{1}{\rho_{v}}\frac{\partial(\rho_{p}w_{T})}{\partial z} \Big\vert^{-1} \approx \Big( \frac{\rho_{v}L_{v}H_{\rm eva}}{F_{p}} \Big) \sim \mathcal{O}(10^{8}\;{\rm s}),
\end{split}
\end{equation}
where $F_{p} \approx 7.5\;{\rm W\;m^{-2}}$ is the planetary heat flux, $L_{v} \approx 2\times 10^{6}\;{\rm J\;kg^{-1}}$ is the specific latent heat, and $\rho_{v} \approx 0.02\;{\rm kg\;m^{-3}}$ is the local density of water vapor if the atmospheric oxygen abundance is three times solar. The diabatic timescale of water vapor has the same magnitude as the characteristic timescale of the hydrological cycle, as suggested by previous studies \citep{gierasch1976jovian,ge2024heat}.

The diabatic changing timescale of buoyancy due to reevaporation is,
\begin{equation}
\begin{split}
    \tau_{\theta_{v}} & = \Big\vert q_{v}\Big(\frac{L_{v}}{c_{p}T} + \frac{\epsilon-1}{\epsilon} \Big) \Big\vert^{-1} \tau_{q_{v}} \sim \mathcal{O}(10^{10}\;{\rm s}),
\end{split}
\end{equation}
where $c_{p}$ is the specific heat capacity at constant pressure, and $\epsilon$ is the molecular weight ratio of water vapor to the hydrogen-helium mixture (see Eq.~\ref{eq:full-thetav} and \ref{eq:tau-thetav} for derivations). $\tau_{\theta_{v}}$ is significantly longer than $\tau_{q_{v}}$ by a factor of $1/q_{v}$, so the buoyancy of air parcels is less sensitive to the temperature and composition change due to reevaporation of precipitation.

Finally, the diabatic changing timescales of PV, $\tau_{\rm PV}$, are dominated by the divergence of the latent heat flux and density change due to reevaporation (i.e., see Eq.~\ref{eq:full-PV}-\ref{eq:tau-diff-heating} for derivations and discussions),
\begin{equation}
\label{eq:tau-PV}
\begin{split}
    \tau_{\rm PV-evap} & \approx \Big\vert \frac{g}{\rho N^{2}}\Big( \frac{L_{v}}{c_{p}T} + \frac{\epsilon-1}{\epsilon} \Big) \frac{F_{p}}{L_{v} H^{2}_{\rm eva}} \Big\vert ^{-1} \\
    & = \Big\vert \frac{gq_{v}}{ N^{2}H_{\rm eva}}\Big( \frac{L_{v}}{c_{p}T} + \frac{\epsilon-1}{\epsilon} \Big) \Big\vert ^{-1} \tau_{q_{v}} \\
    & \sim \mathcal{O}(10^8 \; {\rm s}),
\end{split}
\end{equation}

\begin{equation}
    \tau_{\rm PV-precip} = \Big\vert \frac{1}{\rm PV}\frac{\rm PV}{\rho}\frac{\partial (\rho_{p}w_{T})}{\partial z} \Big\vert^{-1} = \tau_{q_{v}} \sim \mathcal{O}(10^{8}\;{\rm s}).
\end{equation}
which are about the same magnitude and suggest the diabatic changing timescale of PV is on the magnitude of water vapor $\tau_{q_{v}}$. 

A noticeable difference between PV and the other tracers is that the diabatic changing timescale of PV might be relatively short if PV becomes substantially small at nearly neutrally stratified layers close to the LCL or beneath 12 bars. Because PV is nearly zero, the diabatic mixing timescale of PV becomes much shorter than the timescale of water vapor.

Suppose Jupiter's atmospheric water abundance is super solar. In that case, all of these diabatic changing timescales are about one to a few years and, therefore, might be sufficiently longer than the timescale $\tau_{H}$, which is on the magnitude of about ten days. Thus, if Jupiter's atmosphere possesses super-solar water, we may treat water vapor, virtual potential temperature, and PV as quasi-conserved tracers within a few wave cycles. On the other hand, if Jupiter's atmospheric water abundance is reduced to subsolar, then at least $q_{v}$ and PV are not likely to be quasi-conserved due to a relatively more vigorous diabatic mixing.

\section{Conclusions and Discussions}
\label{sec:discussion}

In this study, we suggest that water exhibits a non-uniform distribution in Jupiter's mid-latitudes. Our simulation and analysis indicate that falling precipitation contributes to large-scale vertical depletion of water vapor and the formation of a weakly stratified atmosphere beneath the LCL (Eq.~\ref{eq:eddy-water-flux} and Fig.~\ref{fig:fig2-hydrological-cycle}). We suggest that large-scale nonlinear eddies and waves sustain a meridional dependency of water vapor concentration below the LCL. We propose that PV can be used to diagnose water vapor transport and possibly its distribution, given its similar distribution to water vapor mass fraction (Fig.~\ref{fig:fig1-water}). Further studies that include Lagrangian particles may further accurately constrain the mixing timescale and trajectory \citep[e.g.,][]{kelly1991wintertime}. Our study also extends the application of PV from idealized 2D quasi-geostrophic or shallow water models to a 3D setting on Jupiter, relating the density stratification to water vapor abundance (Eq.~\ref{eq:stratification}).

The non-uniform distribution of water vapor necessitates distinct forcing mechanisms on Earth and Jupiter. On Earth, insolation --- an external forcing --- plays a crucial role. At small scales, sea surface temperature (SST) influences moist convection, while at larger scales, baroclinic eddies driven by differential solar heating transport water vapor across latitudes \citep{emanuel1994large,held2006robust,schneider2010water,payne2020responses}. In contrast, Jupiter's weather layer may be less responsive to solar heating due to its longer radiative timescale. Our simulation indicates that the meridional gradient of water vapor on Jupiter can emerge due to the $\beta$ effect under spatially uniform cooling and heating without requiring differential solar heating. The background stratification set up by the moist system is one of the keys that feed back to the water distribution. Nevertheless, our study suggests that the two systems share a very similar baseline, whereby the moist system interacts with the large-scale dynamics by setting the statistical equilibrium state \citep{emanuel1994large}, including stratification and vertical mixing timescales \citep{o2011effective}.

Although the mid-latitude thermal and compositional structure suggested in this study may relate to the `jovicline' discovered by the Juno spacecraft (Fig.~\ref{fig:fig1-water} and Movie S1), there are still some open questions about the water system and moist convection on Jupiter. Lightning is often considered an indicator of the strength and frequency of localized storms and moist convection, with its occurrence rate exhibiting a meridional dependence \citep{little1999galileo,brown2018prevalent}. Although our large-scale simulation at current spatial resolutions can capture localized water vapor and energy flux, it struggles to resolve the full range of relevant length scales, spanning more than six orders of magnitude, from planetary-scale waves to lightning phenomena. Localized large-eddy simulations may help establish the relation between convective available potential energy (CAPE) and velocity shear for lightning. These relationships could then be applied to explain the observed variation in lightning frequency across latitudes. Another key consideration is whether slow diabatic mixing within the water weather layer can account for the observed concentrations of disequilibrium species, such as CO, $\rm PH_{3}$, $\rm AsH_{3}$, and $\rm GeH_{4}$, especially the M-band observations have detected a possible latitudinal dependency of $\rm AsH_{3}$ and $\rm GeH_{4}$ \citep{noll1990abundance,giles2017latitudinal,grassi2020spatial,bjoraker2022spatial,cavalie2023subsolar}. Tropospheric CO and $\rm PH_{3}$ abundance might be reproduced with three times solar water vapor under weak vertical mixing \citep{wang2016modeling,cavalie2023subsolar,hyder2024supersolar}, while whether $\rm GeH_{4}$ and $\rm AsH_{3}$ would reconcile with the super solar water remains unclear. However, the kinetics of As and Ge chemical networks in giant planet atmospheres remain largely uncertain to accurately constrain Jupiter's vertical mixing due to the scarcity of experimental data \citep{wang2016modeling,hyder2024supersolar}.

Future studies can build upon our findings to unveil the global distribution of water vapor and other chemical species beyond Jupiter's mid-latitudes. While our study focused on mid-latitude water vapor distribution, expanding this analysis to a global scale --- particularly the equatorial and polar regions --- will provide a more comprehensive understanding of Jupiter's atmospheric composition. The equatorial zone, with its near-zero or negative PV and superrotation driven by angular momentum convergence \citep{schneider2009formation,liu2010mechanisms}, may exhibit unique characteristics that warrant further investigation.

Additionally, future research will explore the formation and stability of jets, which play a crucial role in shaping the staircase structures observed in Fig.~\ref{fig:fig1-water} and Movie S1. Another avenue for future exploration is refining the representation of atmospheric forcing in models. While our study did not include differential solar heating, radiative transfer calculations indicate that most insolation is either reflected by upper clouds or absorbed above the 2-bar level, with a comparable meridional heat flux to the planetary heat flux \citep{li2018high,guerlet2020radiative}. Given Jupiter's size and relatively weak solar heating, the effect of differential heating on PV may be minor; however, previous studies have suggested that its influence on deep atmospheric temperature, jet formation, and density structure remains \citep{ingersoll1978solar,schneider2009formation}.

Further studies can also reveal a more realistic structure of water vapor depletion by advancing our understanding of cloud microphysics and precipitation. While we suggest that water depletion extends to the level where precipitation fully evaporates, detailed microphysics studies on precipitation reevaporation and associated small-scale turbulence can provide more precise estimates of how deep condensates fall \citep{guillot2020stormsI,guillot2020stormsII,markham2023rainy}. Lastly, future research can provide a more comprehensive understanding of the differential heating induced by solar heating and cloud-radiative feedback, tracer mixing, jet formation, and the role of baroclinic modes near prograde jets. Prograde jets may be key to understanding water vapor transport in the mid-latitudes, where sharp meridional PV gradients and sloped mixing near prograde jets (Fig.~\ref{fig:Fig4-correlation} and \textit{SI Appendix} Figs.~3 and 4). Addressing these questions will further illuminate the complex interplay of atmospheric dynamics and chemical tracers on Jupiter and other giant planets.

\section*{acknowledgement}
We thank the reviewers for constructive comments and suggestions. H.G. is supported by NASA Earth and Space Science Fellowship 80NSSC18K1268, the dissertation quarter fellowship from UC Santa Cruz, and 51 Pegasi b fellowship (Grant \#2023-4466) from the Heising-Simons Foundation. C.L. was supported by the Heising-Simons Foundation and NASA's New Frontier Data Analysis Program grant 80NSSC23K0790. X.Z. acknowledges support from the National Science Foundation grant AST2307463, NASA Exoplanet Research grant 80NSSC22K0236, and the NASA Interdisciplinary Consortia for Astrobiology Research (ICAR) grant 80NSSC21K0597. A.P.I. and S.C. are supported by the NASA grant 80NSSC20K0555 and a contract with the Juno mission, which is administrated for NASA by the Southwest Research Institute. H.G. acknowledges supercomputer Lux at the University of California, Santa Cruz, funded by NSF MRI grant AST 1828315. Resources supporting this work were provided by the NASA High-End Computing (HEC) Program through the NASA Advanced Supercomputing (NAS) Division at Ames Research Center.

\appendix

\section{Governing Equations}
\label{app:gov-eqs}

The mass continuity equations of the dry air, water vapor, water cloud, and water precipitation can be written as,

\begin{equation}
\label{eq:dry-continuity}
    \frac{\partial \rho_{d}}{\partial t} + \nabla\cdot(\rho_{d}\boldsymbol{u}) = 0,
\end{equation}

\begin{equation}
\label{eq:vapor-continuity}
    \frac{\partial \rho_{v}}{\partial t} + \nabla\cdot(\rho_{v}\boldsymbol{u}) = \mathcal{R}_{\rm evap} - \mathcal{R}_{\rm cond},
\end{equation}

\begin{equation}
\label{eq:cloud-continuity}
    \frac{\partial \rho_{c}}{\partial t} + \nabla\cdot(\rho_{c}\boldsymbol{u}) = \mathcal{R}_{\rm cond} - \mathcal{R}_{\rm auto},
\end{equation}

\begin{equation}
\label{eq:precip-continuity}
    \frac{\partial \rho_{p}}{\partial t} + \nabla\cdot(\rho_{p}\boldsymbol{u}) + \nabla\cdot(\rho_{p}\boldsymbol{V_{T}}) = \mathcal{R}_{\rm auto} - \mathcal{R}_{\rm evap},
\end{equation}
where $\rho_{d}$ is the dry air density, $\boldsymbol{u} = (u,v,w)^T$ is the 3D velocity field of air parcels under the Eulerian frame, $\rho_{v}$ is the water vapor density, $\rho_{c}$ is water cloud density, $\rho_{p}$ is the water precipitation density, $\boldsymbol{V_{T}} = (0,0,w_{T})^T$ is the 3D terminal velocity vector of precipitation, $\mathcal{R}_{i}$ is the rate in mass due to phase transition where $i$ stands for reevaporation, condensation, auto-conversion (e.g., collision-coalescence). We implicitly assume that the cloud particles, which have sub-micron sizes with a small Stokes number, are perfectly coupled with the background flow. On the other hand, precipitating condensates (i.e., larger condensates) are decoupled from the background flow with a larger Stokes number and reach their terminal velocity instantaneously.

SNAP solves the full conservative form of the momentum equations. The primitive form of inviscid momentum equations from \citep{li2019simulating} for deriving PV is,
\begin{equation}
    \frac{\partial \boldsymbol{u}}{\partial t} + (2\boldsymbol{\Omega}+\nabla\times\boldsymbol{u})\times\boldsymbol{u} = -\frac{1}{\rho}\nabla p - \nabla(gz+\frac{1}{2}\boldsymbol{u}\cdot\boldsymbol{u}) + \boldsymbol{F}_\text{drag},
\end{equation}
where $\boldsymbol{\Omega}$ is the angular velocity of the planetary rotation with shallow approximation by neglecting its impact on the vertical velocity; $gz$ is the gravitational potential assuming the gravity is a constant and the reference height, $z = 0$ m, is at the 1 bar level initially; $p$ is the atmospheric pressure; and $\boldsymbol{F}_\text{drag}$ is the specific drag force \citep[Eq.~4.56 in][]{vallis2017atmospheric}. 

The total energy equation reads as,
\begin{equation}
    \frac{\partial E}{\partial t} + \nabla\cdot \Big[ u(E + p) \Big] = -\rho g w - \rho_{p}gw_{T}-\frac{\partial(E_{p}w_{T})}{\partial z},
\end{equation}
where $E = \sum_{i}\rho_{i}c_{v,i}T + \rho(u^2 + v^2 + w^2)/2 + \sum_{i\neq d}\rho_{i}\mu_{i}$ is the total energy with $c_{v,i}$ stands for the isochoric heat capacity of species $i \in \{ d,v,c,p \}$, $\mu_{i}$ stands for the chemical potential of species $i$ to update the latent heat from the Clausius-Clapeyron relation, $E_{p}$ stands for the total energy of precipitation $E_{p} = \rho_{p}[c_{v,p}T + (u^2 + v^2 + w^2)/2 + \mu_{p}]$, where we ignore the kinetic energy arise from the falling velocity to simplify the analysis as it is usually negligible compared to the rest terms. See Eq.~8 to 16 in \citep{li2019simulating} for the detailed description of the energy equation used in the model.

\section{Model Setup}
\label{app:model-setup}

Here, we provide more details about the numerical experiments so that one can reproduce our results. The computational cost of a single case is comparable to that of climate modeling, resolving a few hydrological cycles, which requires approximately 1,000 Intel Xeon E5-2680v2 processors (2.8 GHz) to run for several months to achieve the statistical steady state (\textit{SI Appendix} Fig.~1 and \textit{SI Appendix} Movie~2). Limited by computational resources, we perform a single case simulation with $\rm 3 \times$ solar water and explore the significance of metallicity, planetary heat flux, and cloud microphysics processes through analytical estimations. To improve the computational efficiency, we enable the vertically implicit scheme to suppress the vertically propagating acoustic waves \citep{ge2020global}. We fix the CFL number as one. The corresponding time step is approximately 80 seconds. The planetary rotational frequency, $\Omega$, is fixed as $\rm 1.76\times 10^{-4} \; s^{-1}$. The meridional gradient of the Coriolis parameter (i.e., planetary vorticity), $\beta_{0}$, is fixed as a constant for the $\beta$-plane simulation, $\beta_{0} = 2\Omega/a \approx 4.92\times 10^{-12} \; {\rm m^{-1}\;s^{-1}} $, where $a$ is Jupiter's planetary radii. Hence, the Coriolis parameter, $f$, is linear in latitude and calculated by $f = 2\Omega\sin{\phi_{0}} + \beta_{0} y$, where $\phi_{0} = 20^{\circ}$ N is the southern boundary of the domain and $y$ is the meridional distance from the southern boundary. We adopt reflecting and free-slip boundary conditions in the meridional and vertical directions, but periodic boundary conditions in the longitudinal direction. We implement horizontally uniform heating at the bottom of the domain and uniform body cooling above the 1 bar level. The heating and cooling fluxes are fixed to 7.5 $\rm W\;m^{-2}$ to maintain the energy budget of the entire system. The simulation is initialized with 3 times solar water vapor. The water abundance is relaxed to the initial condition at the lower boundary to mimic the recycling and replenishment of moisture in the deep convective layer. The relaxation timescale is chosen as $10^4$ s to match a general convective timescale in the atmosphere at about 100 bars. The timescale is inferred from the mixing length theory. The dry part of the atmosphere has a molecular weight of 2.2 $\rm g\;mol^{-1}$. Our model employs the finite volume method (FVM) framework, as it has a better performance in conserving tracers. The vertical implicit scheme in our model is also carefully designed for conservation laws (see benchmark tests in \citep{li2019simulating,ge2020global}). The good numerical stability of our model also allows us to adopt a low Mach number Riemann solver, thereby better resolving various gradients, including the beta effect that is important for PV. Movies 1 and 2 also demonstrate the performance of PV conservation in a timescale of tens of days.

\section{Mass Conservation of Total Water and Turbulent Transport of Water Vapor}
\label{sec:ref-eddy-vapor-transport}

One can derive the continuity equation of total water content, $\rho_{w} = \rho_{v} + \rho_{c} + \rho_{p}$ from Eq.~\ref{eq:vapor-continuity}-\ref{eq:precip-continuity},
\begin{equation}
\label{eq:total-water-continuity}
    \frac{\partial \rho_{w}}{\partial t} + \nabla\cdot(\rho_{w}\boldsymbol{u}) + \nabla\cdot(\rho_{p}\boldsymbol{V_{T}}) = 0.
\end{equation}
Similarly, the conservation of all materials (i.e., the dry component, water vapor, and water condensates) can be written as,
\begin{equation}
\label{eq:total-atm-continuity}
    \frac{\partial \rho}{\partial t} + \nabla\cdot(\rho\boldsymbol{u}) + \nabla\cdot(\rho_{p}\boldsymbol{V_{T}}) = 0.
\end{equation}

We then derive the eddy transport equation of total water. We will show that, to satisfy the conservation of total water, there must be an upward eddy vapor flux to balance the downward sedimentation flux,
\begin{equation}
\label{eq:water-eddy-transport}
    \overline{q_{v}'w'} = -\dfrac{(1 - \overline{q_{v}})}{\overline{\rho}}\overline{\rho_{p}w_{T}},
\end{equation}
where $q_{v} = \rho_{v}/\rho$ is the local mass fraction of water vapor. We introduce the horizontal mean of physical quantity $A$ over a sufficiently long time at the quasi-steady state, so we have $A(x,y,z,t) = \overline{A}(z) + A'(x,y,z,t)$. If we take an average on Eq.~\ref{eq:total-water-continuity} and Eq.~\ref{eq:total-atm-continuity}, the time derivative and the horizontal flux divergence would vanish. Then, we have the flux form of the mass continuity equations,
\begin{equation}
\label{eq:water-flux}
    \overline{\rho_{v}w} = -\overline{\rho_{p}w_{T}} - \overline{\rho_{p}w} \approx -\overline{\rho_{p}w_{T}},
\end{equation}

\begin{equation}
\label{eq:total-flux}
    \overline{\rho w} = -\overline{\rho_{p}w_{T}}.
\end{equation}
Because the gravitational settling dominates the precipitation flux, we neglect the eddy precipitation flux due to the background flow, $\overline{\rho_{p}w}$. Recalling the Reynold's decomposition, we have,
\begin{equation}
\label{eq:rho-w}
    \overline{\rho w} = \overline{\rho'w'} + \bar{\rho} \bar{w} = \overline{\rho'w'},
\end{equation}

\begin{equation}
\label{eq:rho_v-w}
    \overline{\rho_{v}w} = \overline{(\overline{\rho}+\rho')(\overline{q_{v}}+q_{v}')(\overline{w}+w')} \approx \overline{q_{v}}\overline{\rho' w'} + \overline{\rho}\overline{q_{v}'w'}
\end{equation}
by neglecting the high-order term $\overline{\rho' q_{v}' w'}$, where $\overline{w} = 0$ (the simulated weather layer $\overline{w}$ is on the magnitude of $\rm 10^{-9}$ $\rm m\;s^{-1}$) and $\overline{\rho} \gg \rho'$ because of the hydrostatic approximation for the reference atmosphere at the averaged steady state. It is noteworthy that $q_{v}' \ll \overline{q_{v}}$ is not necessary for the derivation. Simulation shows that at certain levels, it is possible that $\mathcal{O}(q_{v}') \sim \mathcal{O}(\overline{q_{v}})$ (e.g., Fig.~\ref{fig:fig1-water}A). Substituting Eq.~\ref{eq:rho_v-w} and Eq.~\ref{eq:rho-w} into Eq.~\ref{eq:water-flux} and Eq.~\ref{eq:total-flux}, we have,
\begin{equation}
\label{app-eq:water-eddy-transport}
    \overline{q_{v}'w'} = -\dfrac{(1 - \overline{q_{v}})}{\overline{\rho}}\overline{\rho_{p}w_{T}},
\end{equation}
which is Eq.~\ref{eq:eddy-water-flux} and~\ref{eq:water-eddy-transport}.

If the deep reservoir is homogeneous, the only way to create a concentration anomaly, $q_{v}'$, is the reevaporation of precipitation since advection cannot generate a concentration anomaly in a well-mixed environment $Dq_{v}/Dt = 0$. However, the reevaporation of precipitation always creates a cold pool with moisture-enriched anomalies. It must have negative buoyancy that tends to bring positive $q_{v}$ downward. The favor of the anticorrelation between $q_{v}'$ and $w'$ causes a downward eddy vapor flux, $\overline{q_{v}'w'} < 0$. Therefore, because the RHS of Eq.~\ref{eq:eddy-water-flux} is always positive, a well-mixed vapor reservoir is impossible to balance the sedimentation flux and close the steady-state hydrological cycle with a favored downward vapor flux, $\overline{q_{v}'w'} < 0$.

The turbulent transport of water vapor is slightly different from the mass flux analysis in \citep{guillot2020stormsII} since \citep{guillot2020stormsII} did not distinguish the eddy water vapor flux generated by concentration anomaly $q_{v}'$ from the flux generated by the density anomaly $\rho'$. The down-gradient transport is only necessary if $q_{v} \ll 1$. Otherwise, the water vapor transport falls into a different regime if the atmosphere is mostly water vapor ($q_{v} \sim 1$). In pure water worlds, the eddy water vapor flux is generated by $\rho'$, and a background water vapor gradient is unnecessary.

\section{Stratification due to Precipitation --- Buoyancy Flux Budget}
\label{app:stratification}

Falling precipitation in the atmosphere performs as sinking rocks in pools. The gravitational potential energy of precipitation could convert into three types of energy of the background: (I) the turbulent kinetic energy (TKE) associated with waves; (II) the gravitational potential of the gaseous atmosphere that makes the neutrally or stratified atmosphere more buoyant; (III) internal energy due to friction. Internal energy change is usually negligible, while the flux Richardson number determines the relative significance between TKE and total net buoyancy flux. Here, we first neglect the TKE change and assume the precipitation effect entirely contributes to the buoyancy flux of the gaseous atmosphere as a first-order understanding. The net buoyancy flux of the gaseous atmosphere is $g\overline{w'\theta_{v}'}/\theta_{v}$, and the buoyancy flux of the precipitation is $\overline{q_{p}gw_{T}}$. By approximating the buoyancy flux as an eddy diffusive process with the eddy diffusivity $\mathcal{K}_\text{mom}$, then we have,
\begin{equation}
\label{app-eq:buoyancy-flx}
    \frac{g}{\theta_{v}}\overline{w'\theta_{v}'} = -\frac{g}{\theta_{v}}\mathcal{K}_\text{mom} \overline{\frac{\partial\theta_{v}}{\partial z}} = g\overline{q_{p}w_{T}}
\end{equation}
On the other hand, the precipitation flux is balanced by the eddy water vapor flux if $q_{v} \ll 1$ according to Eq.~\ref{eq:eddy-water-flux},
\begin{equation}
\label{app-eq:vapor-flx}
     \overline{q_{p}w_{T}} \approx -\overline{q_{v}'w'} = \mathcal{K}_{q_{v}}\overline{\frac{\partial q_{v}}{\partial z}}.
\end{equation}
Assuming the eddy diffusivity of water vapor is the same as the eddy diffusivity of the momentum flux, $\mathcal{K}_{q_{v}} = \mathcal{K}_\text{mom}$, in which the eddy fluxes are linear to their dimensionless gradients, we have the buoyancy frequency
\begin{equation}
\label{app-eq:stratification}
    N^{2} \approx \frac{g}{\theta_{v}} \overline{\frac{\partial \theta_{v}}{\partial z}} \approx -g\overline{\frac{\partial q_{v}}{\partial z}},
\end{equation}
which is Eq.~\ref{eq:stratification}. The assumption of approximating the eddy momentum and vapor flux by eddy diffusive processes is only purposed to provide a qualitative understanding of the simulation result. The more realistic non-linear transport of such fluxes shall be numerically integrated by nonhydrostatic models.

\section{Potential Vorticity in Giant Planet Weather Layers}
\label{app:define-PV}

Here, we derive the potential vorticity equation for the weather layers of giant planets. We first prepare continuity equations. The mass continuity equation of a precipitating atmosphere could be written as,
\begin{equation}
    \frac{\partial \rho}{\partial t} + \nabla \cdot (\rho \boldsymbol{u}) + \nabla\cdot(\rho_{p}\boldsymbol{V_{T}}) = 0,
\end{equation}
The primitive momentum equations in a Cartesian beta-plane box are,
\begin{equation}
    \frac{\partial \boldsymbol{u}}{\partial t} + (2\boldsymbol{\Omega}+\nabla\times\boldsymbol{u})\times\boldsymbol{u} = -\frac{1}{\rho}\nabla p - \frac{1}{2}\nabla(gz+\boldsymbol{u}\cdot\boldsymbol{u}) + \boldsymbol{F}_\text{drag},
\end{equation}
where $\boldsymbol{\Omega}$ is the angular velocity of the planetary rotation; $gz$ is the gravitational potential assuming the gravity is a constant and the reference height, $z = 0$ m, is at the 1 bar level; and $p$ is the atmospheric pressure \citep[Eq.~4.56 in][]{vallis2017atmospheric}. We ignore the vertical acceleration due to the aerodynamic drag from the settling of precipitation. We approximate the density of all gaseous species as the total density, $\rho$, because the mass fraction of total condensates, $\rho_{c}$ and $\rho_{p}$, is much smaller than one. In simulations, the mass fraction of total condensates is on the magnitude of $10^{-5}$ to $10^{-7}$ $\rm kg\;kg^{-1}$ and usually much smaller below the LCL (Fig.~\ref{fig:fig2-hydrological-cycle}).

In an atmosphere with changing composition and significant molecular weight differences between different species, one should use virtual potential temperature instead of potential temperature to measure the stratification in PV. The equivalency of isentropes and isopycnals no longer holds in this scenario, as air parcels with the same entropy may have density differences. The virtual potential temperature is
\begin{equation}
\label{eq:def-thetav}
    \theta_{v} = \theta\Big[1 - q_{v}(1-\frac{1}{\epsilon})\Big] = T\Big(\frac{p_{r}}{p}\Big)^{{R}/{c_{p}}}\Big[1 - q_{v}(1-\frac{1}{\epsilon})\Big].
\end{equation}
where $\theta$ is the potential temperature, $p_{r}$ is the 1-bar reference pressure, and $\epsilon$ is the molecular weight ratio of water vapor to the hydrogen-helium mixture that is about 7.

Combining the mass continuity equation, the equation for the virtual potential temperature $\theta_{v}$, and the curl of the momentum equation (i.e., equation for absolute vorticity),
\begin{equation}
    \frac{\partial(2\boldsymbol{\Omega}+\nabla\times\boldsymbol{u})}{\partial t} + \nabla\times(2\boldsymbol{\Omega}+\nabla\times\boldsymbol{u})\times\boldsymbol{u} = \frac{1}{\rho^2}\nabla\rho\times\nabla p + \nabla\times\boldsymbol{F}_\text{drag},
\end{equation}
we have the definition of PV for giant planet weather layers,
\begin{equation}
    {\rm PV} = \frac{(2\boldsymbol{\Omega}+\nabla\times\boldsymbol{u})\cdot\nabla \theta_{v}}{\rho}.
\end{equation}
The unit of PV is $\rm PVU = 10^{-6}\; K\;m^{-2}\;kg^{-1}\;s^{-1}$. The continuity equation of PV is
\begin{equation}
\label{eq:PV}
\begin{split}
    \frac{D({\rm PV})}{Dt} = & \frac{\nabla \theta_{v}\cdot (\nabla\rho\times\nabla p)}{\rho^3} + \frac{(2\boldsymbol{\Omega}+\nabla\times\boldsymbol{u})\cdot\nabla \dot{\theta_{v}}}{\rho} + \\
    & \frac{{\rm PV}}{\rho}\nabla\cdot(\rho_{p}\boldsymbol{V_{T}}) + \frac{\nabla\theta_{v}}{\rho}\cdot(\nabla\times\boldsymbol{F}_{\rm drag}).
\end{split}
\end{equation}
Because the vertical gradient of the virtual potential temperature, $\partial_{z}\theta_{v}$, is the dominant component of $\nabla \theta_{v}$. The vertical gradient, which determines the background buoyancy frequency, is greater than the meridional gradient, which relates to the thermal wind (e.g., Ri $\gg 1$ in most cases). We further neglect the curl of the drag force due to sedimentation of precipitation, as its gradients are negligible. Hence, we have $\partial_{z} \theta_{v} \gg \partial_{y} \theta_{v} \gg \partial_{x} \theta_{v}$ in the water weather layer. Thus, the equation of PV could be simplified as,
\begin{equation}
\label{eq:simplified-PV}
\begin{split}
    \frac{D({\rm PV})}{Dt} & \approx \dfrac{D}{Dt}\Big(\dfrac{f+\zeta}{\rho}\dfrac{\partial \theta_{v}}{\partial z}\Big) \\
    & \approx \frac{\dfrac{\partial \theta_{v}}{\partial z} \hat{z}\cdot(\nabla\rho\times\nabla p)}{\rho^3} + \frac{(f+\zeta) \dfrac{\partial\dot{\theta_{v}}}{\partial z}}{\rho} + \frac{\rm PV}{\rho}\frac{\partial(\rho_{p}w_{T})}{\partial z}.
\end{split}
\end{equation}

The first term on the right-hand side (RHS) of the Eq.~\ref{eq:PV} represents the PV change due to the baroclinicity. The second term on the RHS of the equation denotes the impact of buoyancy change due to compositional or thermal effects associated with differential heating or cooling (e.g., radiative heating or cooling or chemical potential change). The third term arises from the density change due to the settling and reevaporation of precipitation. 

\section{Diabatic Changing Timescale of $q_{v}$, PV, and $\theta_{v}$}
\label{app:dissipation-timescale}

The diabatic changing timescale of the water vapor mass fraction is relatively simpler to estimate than the other two. The reevaporation of precipitation beneath the LCL is the only source that can change the water vapor mass fraction of an air parcel in the Lagrangian frame,
\begin{equation}
\label{eq:full-qv}
    \frac{Dq_{v}}{Dt} = \dot{q_{v}} = -\frac{1}{\rho}\frac{\partial(\rho_{p}w_{T})}{\partial z}.
\end{equation}
Hence, we can acquire a typical diabatic changing timescale of water vapor, $\tau_{q_{v}}$, 
\begin{equation}
\label{app-eq:tau-qv}
\begin{split}
    \tau_{q_{v}} & = \Big\vert -\frac{1}{\rho q_{v}}\frac{\partial (\rho_{p}w_{T})}{\partial z} \Big\vert^{-1}
    \approx \Big\vert -\frac{1}{\rho_{v}}\frac{\rho_{p}w_{T}}{H_{\rm eva}} \Big\vert^{-1} \\
    & \approx \frac{\rho_{v}H_{\rm eva}L_{v}}{F_{p}},
\end{split}
\end{equation}
where $H_{\rm eva}$ is the distance that precipitation fully evaporates after detaching from the cloud level, and planetary heat flux, $F_{p}$, confines the averaged precipitation flux, $-\rho_{p}w_{T}$, as the averaged latent heat flux $-\rho_{p}w_{T}L_{v}$ does not exceed the planetary heat flux \citep{guillot2020stormsII,ge2024heat}. Here, we use one scale to approximate $H_{\rm eva}$, which is consistent with the simulation, where precipitation mostly evaporates at approximately 15 bars (Fig.~\ref{fig:fig2-hydrological-cycle}). Microphysics studies also suggest that raindrops can fall one scale height after detaching from cloud levels before their complete reevaporation \citep{loftus2021physics}, and hailstones may fall further but cannot reach more than a few scale heights \citep{guillot2020stormsI}. Thus, the diabatic changing timescale of water vapor is on the magnitude of a few years, $\tau_{q_{v}} \sim \mathcal{O}(10^{8}\;{\rm s})$. The most uncertain parameter for the water system on Jupiter is probably the reevaporation distance, $H_{\rm eva}$. Hypothetically, if precipitation is mostly snowflakes, $H_{\rm eva}$ might be one order of magnitude shorter than one scale height. Nevertheless, the corresponding `dissipation' timescale is still long enough for water vapor to be homogenized on constant PV surfaces before being eroded by the reevaporation of condensates.

Eq.~\ref{eq:tau-qv} also indicates that ammonia is not an active tracer that can persistently affect dynamics and potential vorticity like water. Ammonia vapor has a low abundance and condenses at about 0.7 bars, with a small background atmospheric density. Thus, ammonia is not a quasi-conserved tracer.

The diabatic changing timescale of the virtual potential temperature $\theta_{v}$ is related to $\tau_{q_{v}}$ since the density change of an air parcel is caused by the evaporative cooling and compositional change. Neglecting radiative cooling or heating below the water LCL, the continuity equation of $\theta_{v}$ is,
\begin{equation}
\label{eq:full-thetav}
    \frac{D\theta_{v}}{Dt} \approx -\theta_{v}\Big( \frac{L_{v}}{c_{p}T} + \frac{\epsilon-1}{\epsilon}\Big) \frac{Dq_{v}}{Dt},
\end{equation}
where $c_{p}$ is the heat capacity of the air parcel and $\epsilon$ is the molecular weight ratio of water and the background hydrogen-helium mixture, which is about 7. Here, we also neglect the changing pressure and heat capacity due to reevaporation since they have a minor impact on buoyancy compared to the latent heat effect \citep{li2019simulating}. Then, we have the diabatic changing timescale of virtual potential temperature due to reevaporation,
\begin{equation}
\label{eq:tau-thetav}
\begin{split}
    \tau_{\theta_{v}} & = \Big\vert q_{v}\Big(\frac{L_{v}}{c_{p}T} + \frac{\epsilon-1}{\epsilon} \Big) \Big\vert^{-1} \tau_{q_{v}} \gg \tau_{q_{v}}, \; {\rm if}\;q_{v} \ll 1.
\end{split}
\end{equation}

The diabatic change of PV is the most complicated one since it depends on four different physical processes or their divergences: (I) baroclinicity, (II) buoyancy change due to reevaporation, (III) density change due to precipitation, and (IV) friction, such as the drag force from precipitation. The continuity equation of PV is,

\begin{equation}
\label{eq:full-PV}
\begin{split}
    \frac{D(\rm PV)}{Dt} = & \underbrace{ \frac{\nabla\theta_{v}\cdot(\nabla\rho\times\nabla p)}{\rho^3} }_\text{Baroclinicity} + \underbrace{\frac{(2\boldsymbol{\Omega} + \nabla\times\boldsymbol{u})\cdot\dot{\nabla\theta_{v}}}{\rho}}_\text{buoyancy change} + \\
    & \underbrace{ \frac{\rm PV}{\rho}\nabla\cdot(\rho_{p}\boldsymbol{V_{T}}) }_\text{Precipitation} + \underbrace{\frac{\nabla\theta_{v}}{\rho}\cdot(\nabla\times\boldsymbol{F}_{\rm drag})}_\text{Friction}
\end{split}
\end{equation}
It is difficult to provide an analytical estimation for the baroclinicity in Jupiter's weather layer as the first term on the RHS of Eq.~\ref{eq:full-PV}. However, we could get the magnitude of this term from the simulation. \textit{SI Appendix} Fig.~S6 shows that the magnitude of this term is about $10^{-12}\;{\rm PVU\;s^{-1}}$. The zonal wind in the weather layer is generally barotropic with a very weak baroclinicity (\textit{SI Appendix} Fig.~6C). Hence, giving the PV is on the magnitude of $\mathcal{O}(10^{-2}\;{\rm PVU})$, the timescale associated with baroclinic process is,
\begin{equation}
\label{eq:tau-baroclinic}
    \tau_{\rm baro} = \Big\vert \frac{1}{\rm PV}\frac{\nabla\theta_{v}\cdot(\nabla\rho\times\nabla p)}{\rho^3} \Big\vert^{-1} \sim \mathcal{O}(10^{10} \; {\rm s}).
\end{equation}

The second term on the RHS of Eq.~\ref{eq:full-PV} concerns the differential diabatic heating or cooling. The major source of diabatic heating is evaporative cooling in Jupiter's water weather layer beneath the thick cloud layer, rather than the isolation. The second term could be rewritten as a function of reevaporation in the form of precipitation flux divergence,
\begin{equation}
\begin{split}
    \Big( \frac{L_{v}}{c_{p}T} + \frac{\epsilon-1}{\epsilon} \Big) \frac{2\boldsymbol{\Omega} + \nabla\times\boldsymbol{u}}{\rho} \cdot \nabla \Big[ \frac{\theta_{v}}{\rho} \frac{\partial(\rho_{p}w_{T})}{\partial z} \Big],
\end{split}
\end{equation}
assuming the specific latent heat, $L_{v}$, and the averaged specific heat capacity, $c_{p}$, are constant regardless of any change of temperature or composition. To simplify the estimation, we have further assumed that the horizontal divergence of reevaporation is negligible compared to the vertical divergence due to the large horizontal-vertical aspect ratio of the planet. Then, neglecting the trivial vertical change of $\theta_{v}/\rho$, the diabatic changing timescale of the differential cooling is,
\begin{equation}
\label{eq:tau-diff-heating}
\begin{split}
    \tau_{\rm diff} & \approx \Big\vert \frac{1}{\rm PV}\Big( \frac{L_{v}}{c_{p}T} + \frac{\epsilon-1}{\epsilon} \Big)\frac{(f+\zeta)\theta_{v}}{\rho^{2}}\frac{F_{p}}{L_{v} H^{2}_{\rm eva}} \Big\vert ^{-1} \\
    & \approx \Big\vert \frac{g}{\rho N^{2}}\Big( \frac{L_{v}}{c_{p}T} + \frac{\epsilon-1}{\epsilon} \Big) \frac{F_{p}}{L_{v} H^{2}_{\rm eva}} \Big\vert ^{-1} \\
    & \sim \mathcal{O}(10^8 \; {\rm s}).
\end{split}
\end{equation}
The diabatic changing timescale has the same magnitude as the virtual potential temperature and water vapor mass fraction. Although there is a new parameter --- the square of buoyancy frequency and $\tau_{\rm diff}$ seems more sensitive to the reevaporation distance as $\tau_{\rm diff} \propto H_{\rm eva}^2$, $N^2$ almost cancels $H_{\rm eva}$ because stratification $N^2$ relates to the $H_{\rm eva}$ (Eq.~\ref{app-eq:stratification}). The deeper the precipitation can fall, the deeper the water depletion goes. If we fix the deep water abundance, the stratification is weaker with a larger $H_{\rm eva}$.

The diabatic change of PV due to the reevaporation as the third term on the RHS of Eq.~\ref{eq:full-PV} has the same timescale as the change of water vapor because
\begin{equation}
    \tau_{\rm precip} = \Big\vert \frac{1}{\rm PV}\frac{\rm PV}{\rho}\frac{\partial (\rho_{p}w_{T})}{\partial z} \Big\vert^{-1} = \tau_{q_{v}} \sim \mathcal{O}(10^{8}\;{\rm s}).
\end{equation}
Therefore, the diabatic changing timescale of the PV, $\tau_{\rm PV}$, should be the smallest one among $\tau_{\rm baro}$, $\tau_{\rm diff}$, and $\tau_{\rm precip}$, which is on the magnitude of $10^8$ s.

\section{Data Availability}
The numerical model, software, and code that were used to generate the simulation result, figures, and movies in this paper are posted on Zenodo (DOI:10.5281/zenodo.17102356).

\bibliography{main}{}
\bibliographystyle{aasjournalv7}



\end{CJK*}
\end{document}